\documentclass[main]{aa}  
\usepackage{graphicx}
\usepackage{txfonts}
\usepackage{amsmath}
\usepackage[utf8]{inputenc}
\usepackage{kotex}
\usepackage{xcolor}
\usepackage{verbatim}

\bibpunct{(}{)}{;}{a}{}{,} 

\def\apjl{{ApJL}}
\def\apjs{{ApJS}}

\def\lax{{$\mathrel{\hbox{\rlap{\hbox{\lower4pt\hbox{$\sim$}}}\hbox{$<$}}}$}}
\def\gax{{$\mathrel{\hbox{\rlap{\hbox{\lower4pt\hbox{$\sim$}}}\hbox{$>$}}}$}}
\def\simlt{\lower.5ex\hbox{$\; \buildrel < \over \sim \;$}}
\def\simgt{\lower.5ex\hbox{$\; \buildrel > \over \sim \;$}}
\def\lum{erg s$^{-1}$}

\def\cm2{cm$^{-2}$}

\def\feii{\ion{Fe}{II}}
\def\mgii{\ion{Mg}{II}}
\def\civ{\ion{C}{IV}}

\def\l5100{$L_{5100}$}
\def\ll5100{$\log L_{\rm 5100}$}

\def\-->{$\rightarrow$}
\def\sf200{${\rm SF}_{\rm 200 days}$}

\begin{document}

\title{Continuum variability in multi-epoch quasar spectra from the Sloan Digital Sky Survey
}

\author{Minjin Kim\inst{1} \and Suyeon Son\inst{2,3} \and Luis C. Ho\inst{2,4}}

\institute{Department of Astronomy, Yonsei University, 50 Yonsei-ro, Seoul 03722, Korea\\
\email{mkim.astro@yonsei.ac.kr}
\and
Kavli Institute for Astronomy and Astrophysics, Peking University, Beijing 100871, China
\and
Department of Astronomy and Atmospheric Sciences, Kyungpook National University, Daegu 41566, Korea
\and
Department of Astronomy, School of Physics, Peking University, Beijing 100871, China
}

\date{Received}

\abstract{
We examine the continuum variability of active galactic nuclei (AGNs) by analyzing the multi-epoch spectroscopic data from the Sloan Digital Sky Survey. To achieve this, we utilized approximately 2 million spectroscopy pairwise combinations observed across different epochs for $\sim90,000$ AGNs. We estimate the ensemble variability structure function (SF) for subsamples categorized by various AGN properties, such as black hole mass, AGN luminosity ($L$), and Eddington ratio, to investigate how AGN variability depends on these parameters. We found that the SFs are strongly correlated with $L$, Eddington ratio, and rest-frame wavelength ($\lambda$). The analysis, with each parameter held fixed, reveals that SFs depend primarily on $L$ and $\lambda$, but not on the Eddington ratio. Consequently, under the assumption that AGNs follow a universal SF, we found that the variability timescale ($\tau$) is proportional to both $L$ and $\lambda$, expressed as $\tau \propto L^{0.62\pm0.07} \lambda^{1.74\pm0.23}$. This result is broadly consistent with predictions from the standard accretion disk model ($\tau \propto L^{0.5} \lambda^{2}$). However, when considering only shorter wavelengths ($\lambda \leq 3050\ \AA$) to minimize contamination from the host galaxy and the Balmer continuum, the power-law index for $\lambda$ drops significantly to $1.12 \pm 0.24$. This value is lower than predicted by approximately $3$–$4\ \sigma$, suggesting that the radial temperature profile may be systematically steeper than that predicted by the standard disk model, although other mechanisms may also contribute to this discrepancy. These findings highlight the power of temporal spectroscopic data in probing AGN variability, as they allow robust estimation of continuum fluxes without interference from strong emission lines.}


\titlerunning{Spectroscopic variability of SDSS QSOs}
\authorrunning{Kim et al.}

\maketitle

\section{Introduction} 
Active galactic nuclei (AGNs) are compact objects in which a substantial amount of gas accretes onto supermassive black holes (BHs) at the centers of galaxies, forming an accretion disk that emits intense radiation, often comparable to or even exceeding the total emission of the host galaxy. AGNs produce powerful, multi-wavelength light because their intricate central regions are governed by a broad range of physical processes spanning multiple emission mechanisms \cite[e.g.,][]{elvis_1994, richards_2006}. In addition, the strength of the radiation from the AGNs fluctuates stochastically on various timescales from days to years \cite[e.g.,][]{ulrich_1997}. Notably, the UV/optical continuum, called the big blue bump, arising from the accretion disk, is inherently variable. This variability is echoed by other reprocessed light, like broad emission lines and the near-infrared continuum \cite[e.g.,][]{peterson_1993,kaspi_2000,koshida_2014,kim_2024}. While the physical cause of the variability is still under debate, various studies have claimed that it may be associated with instabilities within the accretion disk \cite[e.g.,][]{kawaguchi_1998}. According to this scenario, the variability timescale is expected to be correlated with the thermal timescale of the accretion disk \cite[e.g.,][]{czerny_1999,burke_2021,tang_2023, son_2025}. In the standard disk model, the thermal timescale is expected to be associated with the orbital timescale, which is proportional to the square of the wavelength and the square root of AGN luminosity.

The light curve of the UV/optical continuum has often been modeled as a damped random walk (DRW) process (\citealt{kelly_2009, kozlowski_2010a}, but see \citealt{kasliwal_2015}). In this framework, the power spectral density (PSD) is characterized by a power law whose index is inversely proportional to the square of the frequency at higher frequencies. Crucially, the PSD becomes flat at a frequency below a specific break frequency. However, the PSD can be accurately estimated only with multi-epoch photometric data with a regular cadence. For this reason, the structure function (SF) has been alternatively used to study the AGN variability \cite[e.g.,][]{kozlowski_2010a, kozlowski_2017, son_2023b}. Computed as the root mean square of brightness at a given time difference ($\Delta t$), the SF can be easily estimated with sparsely sampled light curves obtained from ground-based telescopes. In the SF, the DRW model can be described by a single power law (i.e., SF $\propto \Delta t$) at a timescale lower than the characteristic timescale ($\tau$), equivalent to the break frequency in the PSD, above which the SF flattens above the characteristic timescale \cite[e.g.,][]{Kozlowski_2010}. 

\begin{figure}[pt]
\centering
\includegraphics[width=0.45\textwidth, page=1]{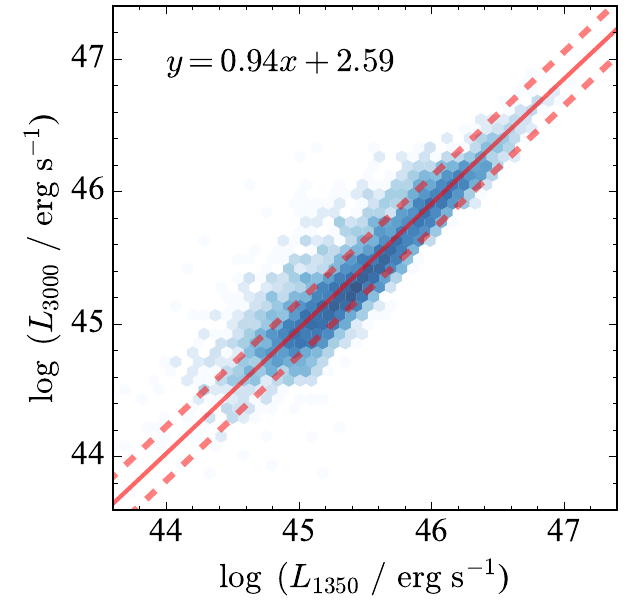}
\caption{
Density distributions of luminosities at 1350 $\AA$ and 3000 $\AA$ for our sample. Red solid and dashed lines denote the best regression and its $1\sigma$ scatter, respectively. The best-fit result displayed in the panel was used to convert between the two luminosities. 
}
\end{figure}

Previous studies based on multi-epoch broadband photometric data widely utilized the SF to characterize AGN variability \cite[e.g.,][]{macleod_2010, kozlowski_2017, li_2018, suberlak_2021, son_2025}. A variety of datasets have been used to infer the variability amplitude from the value of SF at a certain timescale and the characteristic timescale near the knee of the SF \cite[e.g.,][]{sanchez_2017, burke_2023, tang_2023}. Additionally, their dependence on AGN properties [BH mass, AGN luminosity ($L$), Eddington ratio, and rest-frame wavelength ($\lambda$)] was rigorously explored to reveal the mechanism of AGN variability and hence to investigate the structure of the accretion disk. For example, \citet{macleod_2010} estimated the SFs of luminous AGNs in Stripe 82 based on the broadband photometric data of the Sloan Digital Sky Survey (SDSS) over a baseline of $\sim10$ years. They showed that the characteristic timescale ($\tau$) directly estimated from the SFs is correlated with the wavelength and BH mass, but nearly independent of luminosity. Using the PSD of a multi-epoch photometric dataset of $\sim400$ AGNs, spanning a wide range of BH mass ($10^4\leq M_{\rm BH}/M_\odot \leq 10^{10}$), \cite{burke_2023} showed that $\tau$ is mostly governed by the BH mass. Intriguingly, these results are contradictory to the prediction from the standard accretion disk model (\citealt{shakura_1973}; see also \citealt{arevalo_2024, petrecca_2024}).    
More recently, \citet{tang_2023}, using the photometric data from ATLAS (\citealt{tonry_2018}), demonstrated that the characteristic timescale is proportional to $L^{0.539\pm0.004}\lambda^{2.418\pm0.023}$, which is in broad agreement with the theoretical prediction from the standard accretion disk model ($\tau \propto L^{0.5} \lambda^2$). These conflicting results may reflect the potential drawback of the broadband photometric data, which can be contaminated by strong emissions from the broad-line region (BLR) and continuum from neighboring wavelengths owing to the large width of the filter transmission \cite[e.g.,][]{patel_2025}.   

Multi-epoch spectroscopic data offer a better alternative to overcome the limitations of broadband photometric data in characterizing AGN variability. This method enables a more robust estimation of how AGN variability depends on the continuum wavelength. Acquiring time-series spectra of AGNs demands considerable resources, thereby limiting the availability of sufficient datasets for comprehensive studies of AGN variability. Recently, however, several long-term ($\sim$timescales of a few years) monitoring programs, primarily conducted for reverberation mapping (RM) experiments, offer opportunities to study the spectral variability for a large number of AGNs. For instance, \citet{son_2025} used the spectral data of a few hundred luminous AGNs from the SDSS RM project (\citealt{shen_2019, shen_2024}), which spans $\sim7$ years. Their study focused on the SF across various rest-frame wavelengths. Carefully accounting for the uncertainty in the flux calibration of the observed spectra, \citet{son_2025} demonstrated that the variability timescale is proportional to $L^{0.50\pm0.03}\lambda^{1.42\pm0.09}$. This finding suggests that the dependence on AGN luminosity aligns with predictions from the standard accretion disk model. However, the observed dependence on wavelength significantly deviates from the model prediction, implying a temperature profile that is significantly steeper. Although \cite{son_2025} clearly illustrated that the utility of SF derived from the temporal spectral data in exploring the structure of accretion disks, their work was limited to a relatively small sample consisting predominantly of distant high-luminosity AGNs.

Conventional RM projects provide intensive monitoring data for a limited number of AGNs. While valuable, this approach restricts the sample size and the dynamic range of observable AGN properties. This study investigates spectral variability using an alternative method: analyzing multi-epoch spectroscopic observations of a large sample of AGNs from the main SDSS program. This method allows for a significant expansion of the sample size and extends the dynamic ranges of AGN properties, facilitating more comprehensive studies of spectral variability. This work revisits AGN spectral variability, focusing on the SF methodology, by analyzing more than two million spectroscopic pairwise combinations generated from over 90,000 AGNs from the SDSS. Throughout the paper, we adopted the cosmological parameters from the 2018 Planck results ($H_0=67.36 \pm 0.54$ km s$^{-1}$ Mpc$^{-1}$, $\Omega_\Lambda=0.6847 \pm 0.0073$, $\Omega_m = 0.3153 \pm 0.0073$; \citealt{planck_2020}).  

\begin{figure}[t]
\centering
\includegraphics[width=0.45\textwidth, page=1]{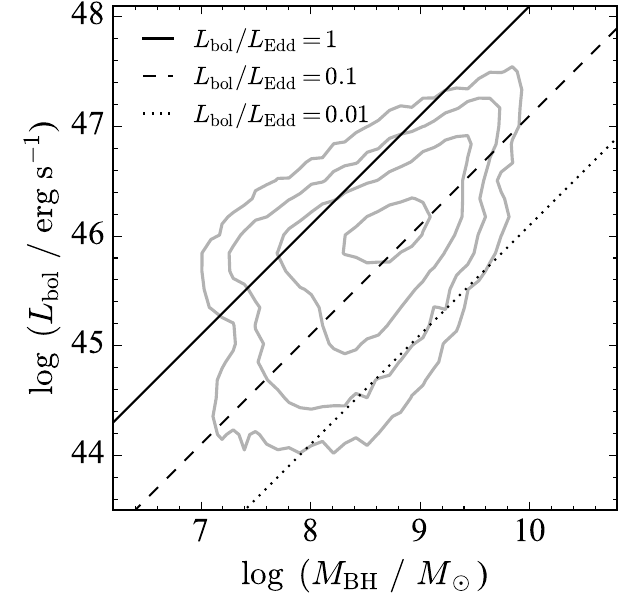}
\caption{
Distributions of BH mass and bolometric luminosity for our sample. The solid, dashed, and dotted lines correspond to Eddington ratios of 1, 0.1, and 0.01, respectively.
}
\end{figure}

\begin{figure}[t]
\centering
\includegraphics[width=0.45\textwidth, page=1]{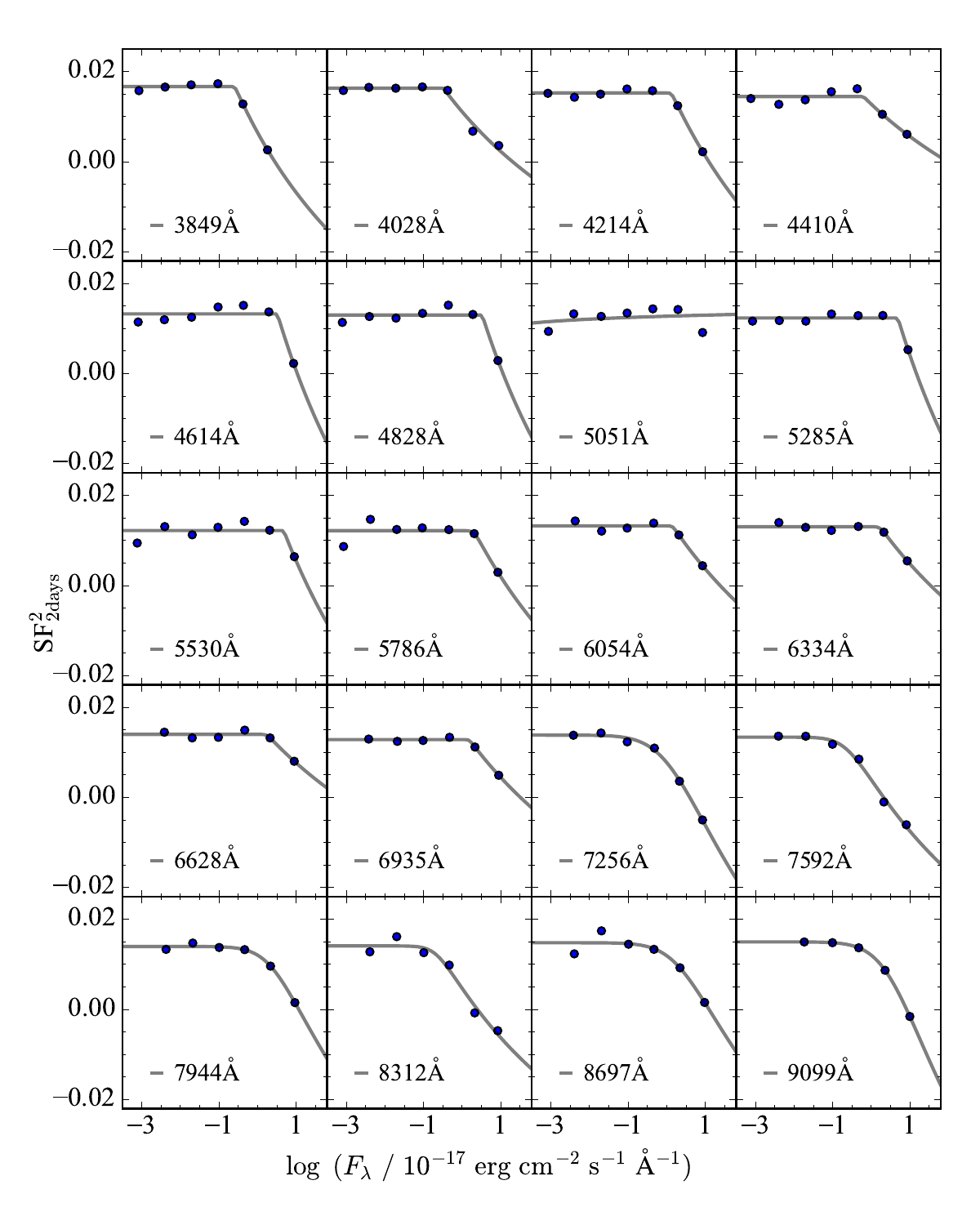}
\caption{
Ensemble SF at $\Delta t < 2$ days (SF$_{\rm 2days}$) plotted against continuum brightness across different wavelengths. The solid line represents the fitting results with a smooth double power-law model.   
}
\end{figure}

\section{Sample and data}
\subsection{Sample selection}
The spectroscopic variability can be investigated in two main ways: (1) through continuous spectroscopic monitoring data of a limited sample of AGNs, primarily conducted for RM analyses (e.g., \citealt{son_2025}), and (2) through multi-epoch spectroscopic datasets, typically two observations for each object, across a large AGN sample. In this study, we adopted the latter method, as a statistically significant amount of data is available in the SDSS. We initially chose the sample from the SDSS QSO catalogs from data release 16 (DR16; \citealt{lyke_2020}) with a selection criterion of \texttt{NSPEC} $\ge 2$, where \texttt{NSPEC} represents the number of duplicate observations. Starting from DR9, the SDSS spectra were acquired using the Baryon Oscillation Spectroscopic Survey (BOSS; \citealt{smee_2013}) spectrograph, which differs from the original SDSS spectrograph in both the fiber aperture and wavelength coverage. To minimize potential systematics in flux calibration due to the instrumental differences, we restrict our analysis to spectra obtained with the BOSS spectrograph from DR9, resulting in a sample of 94,851 objects and 2,233,454 spectroscopic pairwise combinations.  

\begin{figure}[t]
\centering
\includegraphics[width=0.45\textwidth, page=1]{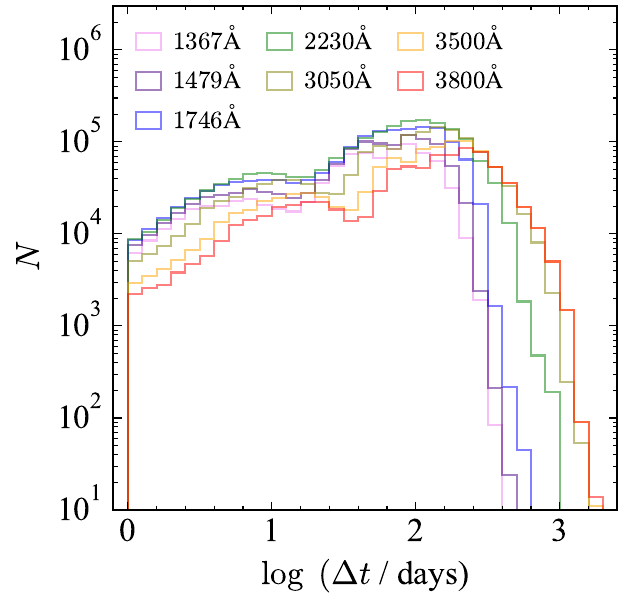}
\caption{
Distribution of observed spectroscopic pairwise combinations at each $\Delta t$ for our sample. Colors denote continuum wavelengths.}
\end{figure}

\subsection{AGN properties}
As the ultimate goal of this study is to investigate how the AGN variability depends on the physical properties of AGNs, it is essential to derive the AGN properties accurately. We directly measured the continuum fluxes ($f_{1367}$ or $f_{3050}$) at 1367 ${\rm \AA}$ or 3050 ${\rm \AA}$ from the mean spectrum of each object in the rest-frame, which is free from strong emission lines. They are corrected for Galactic extinction using $E(B-V)$ values from \cite{schlafly_2011} and the extinction curve of \cite{cardelli_1989} and converted to the conventional luminosities ($L_{1350}$ and $L_{3000}$) at 1350 ${\rm \AA}$ and 3000 ${\rm \AA}$ assuming $f_\lambda \propto \lambda^{-1.56}$ (\citealt{vandenberk_2001}). Figure 1 compares $L_{1350}$ and $L_{3000}$ for objects with both measurements available. As the slope of the correlation clearly deviates from unity, we used an ordinary least-squares linear regression to derive the conversion between the two luminosities: $\log\ (L_{3000} / {\rm erg\  s^{-1}}) = 0.94 \log\ (L_{1350} / {\rm erg\  s^{-1}})+ 2.59$. Finally, the bolometric luminosity ($L_{\rm bol}$) was computed with a bolometric correction of ($L_{\rm bol}=3.81L_{1350}$) adopted from \cite{richards_2006}. We use $L_{1350}$ whenever possible. If $L_{1350}$ is not covered by the SDSS spectrum, $L_{3000}$ serves as an alternative.    

For the BH mass ($M_{\rm BH}$), we used the measurements from \cite{wu_2022}, based on broad \civ\ $\lambda1549$ and \mgii\ $\lambda2800$ emission from the single-epoch spectra\footnote{The BH masses were scaled to reflect the cosmological parameters adopted in this study.}. Priority was given to the \mgii as \civ-based BH masses can suffer from systematic biases \cite[e.g.,][]{ho_2012b, meji_2016}. Subsequently, the Eddington ratio ($L_{\rm bol}/L_{\rm Edd}$) was computed based on the Eddington luminosity $L_{\rm Edd} = 1.26 \times 10^{38} M_{\rm BH}/{M_\odot}\ {\rm erg\  s^{-1}}$.  Radio-loud AGNs are often associated with extreme variability, making them unsuitable for our analysis and thus needing to be excluded \cite[e.g.,][]{vandenberk_2004, macleod_2010}. To identify radio-loud AGNs, we calculated the radio-to-optical flux ratio determined at 2500 $\AA$ ($f_{2500}$) and 6 cm ($f_{\rm 6\ cm}$; \citealt{kellermann_1989}). We initially measured $f_{2500}$ at the nearest continuum windows (2230, 3050, 3500, or 3800 $\AA$) in the rest-frame, depending on the redshift of the target, and subsequently extrapolated to 2500 $\AA$ by assuming a power-law continuum derived from the SDSS composite spectrum ($f_\lambda \propto \lambda^{-1.56}$; \citealt{vandenberk_2001}). $f_{\rm 6cm}$ was estimated from the 20 cm flux density measurements of radio counterparts matched within 2\arcsec\ from the FIRST survey (\citealt{lyke_2020}). For the flux conversion from 20 cm to 6 cm and $k-$correction, we assumed $f_\nu \propto \nu^{-0.5}$ (\citealt{kellermann_1989}; but see \citealt{ho_2001}). Objects with a radio loudness ($R=f_{6cm}/f_{2500}$) greater than 10 (2954 objects) were classified as radio-loud objects and discarded for further analysis.
The distributions of BH mass, bolometric luminosity, and Eddington ratio of the final sample of 91,897 sources are shown in Figure 2.

\begin{figure*}[t]
\centering
\includegraphics[width=0.33\textwidth]{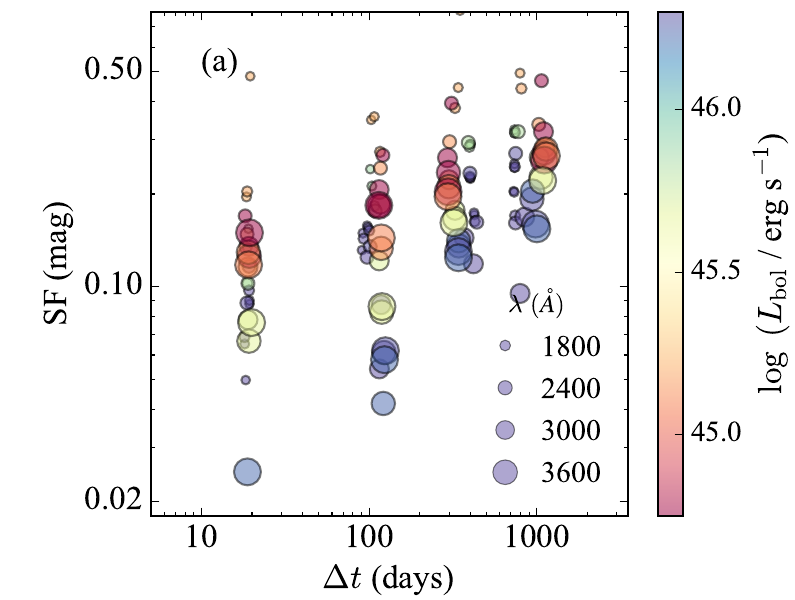}
\includegraphics[width=0.33\textwidth]{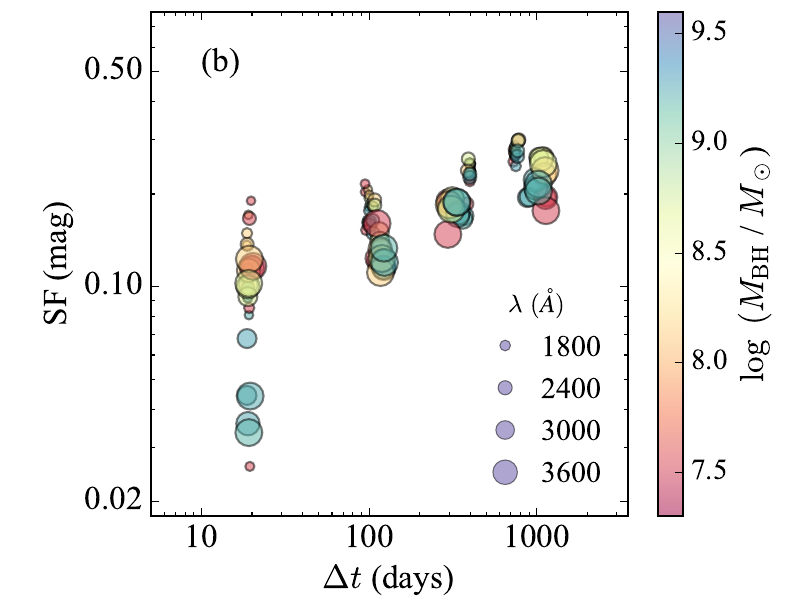}
\includegraphics[width=0.33\textwidth]{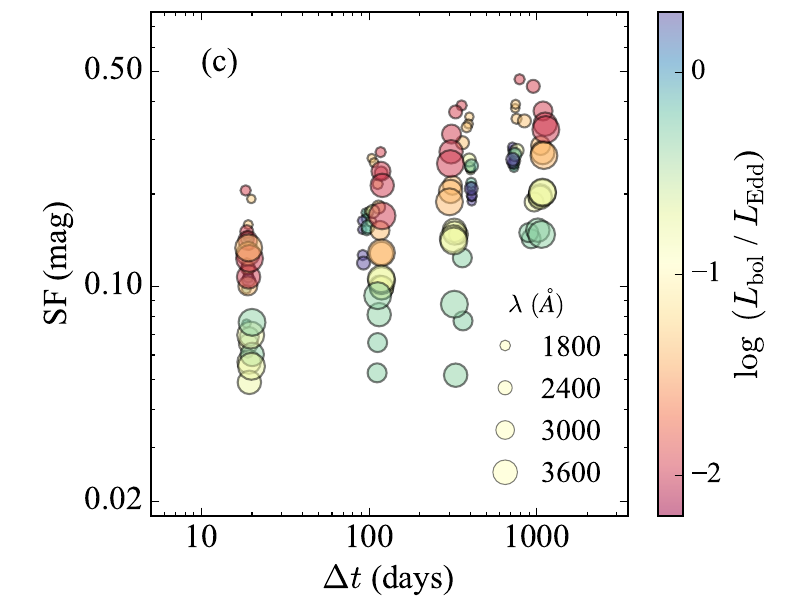}
\caption{
Ensemble SFs for subgroups categorized according to bolometric luminosity ($a$), BH mass ($b$), and Eddington ratio ($c$). The size of the symbols denotes the continuum wavelength. 
\label{fig:fig5}}
\end{figure*}

\subsection{Continuum fluxes and uncertainties}
To minimize the effect of the emission lines, including \feii\ multiples, and starlight from the host galaxy, we estimate the continuum fluxes at 1367, 1479, 1746, 2230, 3050, 3500, and 3800 $\AA$ in the rest-frame (\citealt{son_2025}). At each continuum wavelength, the flux was estimated using the mean flux value, and its uncertainty ($\sigma_{\rm in}$) was calculated as $\sqrt{\frac{\sum_{n=1}^{N}\sigma_i^2}{N}}$, where $N$ is the number of spectral elements within a continuum window\footnote{We investigated the SF with a different width (e.g., 20 \AA) and found that the main results in this paper remained unchanged with the uncertainties} of 40 $\AA$, $\sigma_i$ is the uncertainty associated with each spectral element. However, the flux calibration of the SDSS spectra can introduce systematic uncertainty, which should be considered for accurate SF estimates \cite[e.g.,][]{margala_2016, son_2025}. While, in principle, this uncertainty can be estimated using multi-epoch spectra of standard stars, the standard stars are systematically brighter than our sample, possibly leading to an underestimate of the uncertainty. 

Instead, we utilized the SF of the spectroscopic pairs in the AGN sample with short time intervals ($\Delta t < 2$ days in the observed frame), assuming that the AGN variability on such timescales is negligible. The SF of an observed pair was calculated as 
\begin{equation}
    {\rm SF}^2_{\rm 2\, days} = 
    (m(t) - m(t+\Delta t))^2
    - \sigma_{\rm in}^2(t) - \sigma_{\rm in}^2(t+\Delta t) \quad (\Delta t < {\rm 2 \, days} ), 
\end{equation}
\noindent
where $\Delta t$ is the time lag between the observed pair, $m$ is the observed magnitude at a given continuum wavelength, and $\sigma_{\rm in}$ is the initial uncertainty of magnitude in the single-epoch spectrum. We found that the average SF$^2$ at $\Delta t < 2$ days (SF$^2_{\rm 2\, days}$) is mostly greater than 0 across all the continuum wavelengths, indicating that the original uncertainties ($\sigma_{\rm in}$) were significantly underestimated. Notably, the average SF$^2_{\rm 2days}$ depends on the continuum brightness and is well described by a double power-law model (Fig. 3). We, therefore, incorporate this excess, which depends on the flux density and continuum wavelengths, when estimating the structure function over longer timescales (\S{3.1}). Specifically, the SF$^2_{\rm 2day}$ was modeled using the form: 
\begin{equation}
 {\rm SF}^2_{\rm 2days}= A \left( \frac{f}{f_b} \right) ^ {-\alpha_1} \left\{ \frac{1}{2} \left[1 + \left( \frac{f}{f_b}\right)^{1 / \Delta} \right]\right\}^{(\alpha_1 - \alpha_2) \Delta}, 
\end{equation}
where $f$ is the continuum flux density, $f_b$ is the break flux density, $A$ is the ${\rm SF}^2_{\rm 2days}$ at the low-luminosity end, $\alpha_1$ is the power-law slope at $f_b \gg f$, $\alpha_2$ is the power-law slope when $f \gg f_b$, and $\Delta$ is the smoothness parameter. As $\alpha_1$ was found to be consistent with 0 across wavelengths, we fixed its value to 0 in the fit. As previous studies showed that the flux calibration is highly uncertain at the edges of the BOSS spectrum (\citealt{son_2025}), we restrict our analysis to spectra within the 3820–9150 $\AA$ range to minimize this effect.

\section{Method and results}
\subsection{Analysis of spectral variability with ensemble SF}
To analyze the AGN spectral variability, we utilized the ensemble SF of the spectral dataset. While SFs are typically calculated for individual objects, we instead grouped observed pairs of AGNs with similar physical properties (e.g., BH mass, bolometric luminosity, and Eddington ratio). We then computed the ensemble SF by averaging the individual SFs at each time lag. Therefore, the square of the ensemble SF was calculated as 
\begin{eqnarray}
    {\rm SF}^2(\Delta t) = \frac{1}{N_{\Delta t, {\rm pair}}} \sum_{i=1}^{N_{\Delta t, {\rm pair}}}
    (m(t) - m(t+\Delta t))^2 \nonumber \\
    - \sigma_{\rm in}^2(t) - \sigma_{\rm in}^2(t+\Delta t) - {\rm SF}^2_{\rm 2day}, 
\end{eqnarray}
\noindent
where $N_{\Delta t, {\rm pair}}$ is the number of observed pairwise combinations at the given $\Delta t$ across the AGN subsample, which shares similar AGN properties, and ${\rm SF}^2_{\rm 2day}$ is estimated from the mean magnitude of the observed pairs. It is important to note that $\Delta t$ was expressed in the rest-frame for this calculation. We performed this analysis in each continuum window. The distributions of the number of observed pairs as a function of $\Delta t$ across different continuum windows are shown in Figure 4. This indicates that a sufficient number of pairs exist up to $\Delta t \leq 1000$ days. The extreme AGN variability can occur in a minority of AGNs, manifesting as changing-look AGNs, episodic flares, and tidal disruption events. These phenomena can substantially increase the variability amplitude and thus highly influence the estimation of the ensemble SFs. Therefore, we applied the $3\sigma$ clipping method when averaging the SFs to minimize this effect.

\subsection{Dependence of SF on AGN properties and wavelength}
To explore how the SF varies with AGN properties and continuum wavelengths, we estimated the SFs for subsamples categorized by their AGN properties, such as BH mass, bolometric luminosity, and Eddington ratio. By accounting for the distributions of the physical parameters of the entire sample, we divided the sample into subgroups by five bins covering $10^{44.75}-10^{46.30}\ { \rm erg\ s^{-1}}, 10^{7.3}-10^{9.6}\ M_\odot,$ and $10^{-2.2} - 10^{0.3}$ for the AGN luminosity, BH mass, and Eddington ratio, respectively. On average, there are approximately 2,500 pairwise combinations per bin, given wavelength and $\Delta t$, ranging from 65 to about 19,000.

As illustrated in Figure 5, SFs are highly correlated with the bolometric luminosity and wavelengths. Notably, AGNs with higher luminosity or those at longer wavelengths tend to have lower SFs or longer variability timescales than their lower-luminosity or shorter-wavelength counterparts. 

The standard accretion disk model theoretically predicts that the thermal timescale is intrinsically tied to the orbital timescale. Previous observational studies also proposed that AGN variability timescales are governed by the thermal timescale \cite[e.g.,][]{collier_2001, kelly_2009}. This implies that the observed dependence of SFs on the AGN luminosity and wavelength can be interpreted through variations in the thermal timescale and the radial structure of the accretion disk. Accordingly, this model predicts that the variability timescale is strongly correlated with both AGN luminosity and wavelengths. Motivated by this, we quantified the dependence of the variability timescale on AGN luminosity and wavelength ($\tau \propto L^{\alpha}\lambda^{\beta}$), assuming that all AGNs share a universal ensemble SF when the time lag is normalized by the characteristic timescale ($\tau$) \cite[e.g.,][]{tang_2023, son_2025}. 

We assumed that ${\rm SF} \propto a(\Delta t / \tau)^{b}$, with $\tau$ scaling as $\tau \propto L^{\alpha}\lambda^{\beta}$, where $a$ is the SF at $\Delta t = \tau$, and $b$ is the power-law index for the entire sample. We then performed the fit using the ensemble SFs computed from subsamples divided by AGN luminosity and wavelength. In this calculation, we only include the ensemble SFs derived from more than 500 spectroscopic pairs to minimize a possible systematic error due to small sample sizes (Fig. 6).        
This analysis based on the $\chi^2$ minimization yields that $\tau \propto L^{0.62\pm0.07}\lambda^{1.74\pm0.23}$. At longer wavelengths ($\lambda > 3050\ \AA$), the continuum may be contaminated by the Balmer continuum from the BLR and by stellar light from the host galaxy. In particular, the effect of the Balmer continuum can be complex, as it may be less variable than the underlying continuum and its variability may be delayed due to the extended geometry of the BLR. To mitigate this, we repeated the experiment excluding the wavelengths greater than 3050 $\AA$, resulting in $\tau \propto L^{0.68\pm0.07}\lambda^{1.12\pm0.24}$. Specifically, the power law index ($\beta$) for the wavelength changes dramatically from $\sim1.74$ to $\sim1.12$, suggesting a substantial contribution from either the Balmer continuum or the host light. Additionally, to mitigate the potential impact of outliers in the SFs, we applied a $3\sigma$ clipping during the fitting process. The fitting result remained unchanged within the uncertainties.     

For the Eddington ratio, similar trends were observed in the sense that AGNs with higher Eddington ratios tend to have smaller SFs or longer timescales, possibly due to their correlation with the AGN luminosity (Fig. 5a). With an assumption of the universal ensemble SF, we found $\tau \propto (L_{\rm bol}/L_{\rm Edd})^{0.70\pm0.08}\lambda^{1.99\pm0.24}$ when the fitting was done with the entire wavelength range up to 3800 $\AA$. The fitting results also change significantly to $\tau \propto (L_{\rm bol}/L_{\rm Edd})^{0.67\pm0.08}\lambda^{1.43\pm0.31}$ excluding the wavelength over 3050 $\AA$. However, intriguingly, the dependence of SFs on BH mass is not significant.

\begin{figure}[t]
\centering
\includegraphics[width=0.45\textwidth, page=1]{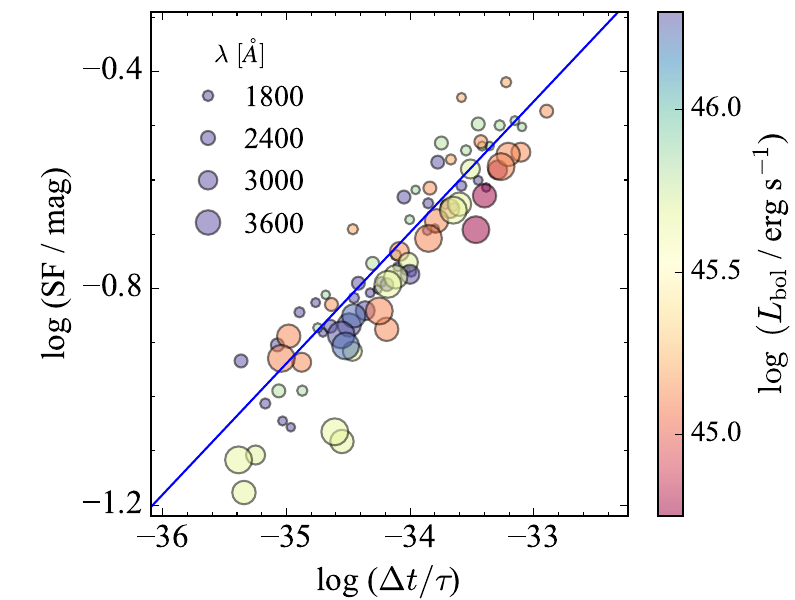}
\caption{
Ensemble SFs with $\Delta t$ normalized by the characteristic timescale ($\tau$). $\tau$, which was estimated assuming that all AGNs share a universal SF, with $\tau$ scaling as $\tau \propto L^{\alpha} \lambda^{\beta}$. }
\end{figure}

\section{Discussion}
\subsection{What is the primary parameter determining the AGN variability?}
Based on the SF analysis for subgroups of the AGN sample, we found that the variability timescales strongly depend on both AGN luminosity and Eddington ratio. As noted in the previous section, the two parameters are closely linked to each other, making it difficult to identify the primary driver of AGN variability. To disentangle this degeneracy, we examined the effect of each parameter by fixing one parameter as a constant while varying the other. First, we selected a subset of AGNs with a narrow range of Eddington ratio ($-1 \le \log\ (L_{\rm bol}/L_{\rm Edd}) \le -0.7$) and recalculated the ensemble SFs for subsamples divided by AGN luminosity. To investigate the dependence of $\tau$ on AGN luminosity, we performed the same analysis under the assumption of a universal SF. This yielded a scaling relation of $\tau \propto L^{0.95\pm0.10}\lambda^{4.15\pm0.36}$. This analysis revealed that the variability timescale retains a significant dependence on AGN luminosity, suggesting that AGN luminosity plays an important role (Fig. 7). However, given the limited sample size\footnote{We performed this experiment using approximately $0.15-0.39$ million pairwise combinations depending on the continuum wavelength.}, the extracted scaling parameters should be interpreted with caution, as they may not reflect true physical relationships. 

On the contrary, we fixed the AGN luminosity within the range of $10^{45.8}-10^{46.2}$ \lum, and assessed the ensemble SFs for subsamples categorized by Eddington ratio. Intriguingly, under these conditions, the dependence on the Eddington ratio vanished, as indicated by the relation $\tau \propto (L_{\rm bol}/L_{\rm Edd})^{-0.01\pm0.06}$ (Fig. 8). This result implies that the Eddington ratio has minimal impact on the variability timescale when AGN luminosity is held nearly constant. 
Taken together, AGN luminosity is the main parameter to govern the variability timescales, rather than the Eddington ratio. 
We note, however, that the marginal correlation between the Eddington ratio and BH mass suggests that secondary effects resulting from this relationship warrant consideration. While this hypothesis could, in principle, be tested by additionally fixing the black hole mass, the limited sample size precludes such an analysis within the scope of this study. Nevertheless, given that the black hole mass appears to exert the least significant influence on the SF across the entire sample (Fig. 5), its impact is likely not dominant.

\subsection{Physical interpretations}
In theoretical studies based on the standard accretion disk framework, the thermal timescale scales as $\tau_{\rm th} \propto M_{\rm BH} R^{3/2}$, where $R$ is the radial distance in the accretion disk. Additionally, the radial temperature profile of the disk follows $T(R) \propto (M_{\rm BH} \dot{M}R^{-3})^{1/4}$, where $\dot{M}$ represents the mass accretion rate. If the AGN luminosity is linearly correlated with $\dot{M}$, the thermal timescale should be proportional to $L^{0.5}\lambda^{2}$ \cite[][]{macleod_2010, tang_2023, son_2025}. Interestingly, this prediction is broadly consistent with our observational findings, which show that the variability timescale, inferred from the SFs, is primarily dependent on the AGN luminosity and wavelength. 

Quantitatively, however, the best-fit power indices derived from our analysis ($\alpha=0.62\pm0.07$ and $\beta=1.74\pm0.23$) exhibit marginal deviations from the theoretical prediction within a $2\sigma$ uncertainty. We note that the deviation in $\beta$ dramatically increases when the wavelengths longer than $3050$ $\AA$, where substantial contamination from the Balmer continuum and stellar light is expected, are excluded from the fit ($\beta = 1.12\pm 0.24$). The fitted value of $\beta$ falls below the theoretical prediction with a significance of $\sim 4\sigma$. This result underscores the critical role of accurately accounting for host galaxy starlight and Balmer continuum contamination in studies of AGN variability. 

Interestingly, the disagreement in $\beta$ with the standard disk model has also been reported in previous observational studies (e.g., \citealt{son_2025}, but see also \citealt{tang_2023}). $\beta$ is directly linked to the radial temperature profile under the assumption that the flux at a given wavelength primarily originates from the radius where the local temperature corresponds to that wavelength via Wien's displacement law ($\lambda \propto 1/T$). The significantly low value of $\beta$ suggests that the radial temperature profile in the accretion disk may be steeper than predicted by the standard disk model \cite[e.g.,][]{son_2025}. This effect can result in a harder spectral energy distribution (SED) of the UV continuum, which is typically contrary to previous findings that report a softer-than-expected UV SED in AGNs \citep[e.g.,][]{zheng_1997, shang_2005}. The larger disk size, relative to predictions from the standard accretion disk model, can also be attributed to a shallower temperature profile \citep[e.g.,][]{sun_2019}. Conversely, an intrinsically harder UV continuum is required to explain the strength of broad emission lines in luminous AGNs \citep[e.g.,][]{lawrence_2012}, which is consistent with a steeper temperature profile.

Alternatively, our results are consistent with predictions from the Corona-Heated Accretion-disk Reprocessing (CHAR) model, in which a steep temperature profile can be reproduced if the flux at a given wavelength is significantly contaminated by emission from neighboring radii \citep{sun_2020}. Within this framework, \citet{zhou_2024} demonstrated that the characteristic timescale depends on AGN luminosity and wavelength, with power-law indices of $\alpha = 0.65$ and $\beta = 1.19$, which are in excellent agreement with our results. The results of \citet{zhou_2024} were derived from simulated light curves constructed using the AGN parameters of QSOs in Stripe 82, as adopted from \citet{stone_2022}.

\begin{figure}[t]
\centering
\includegraphics[width=0.45\textwidth, page=1]{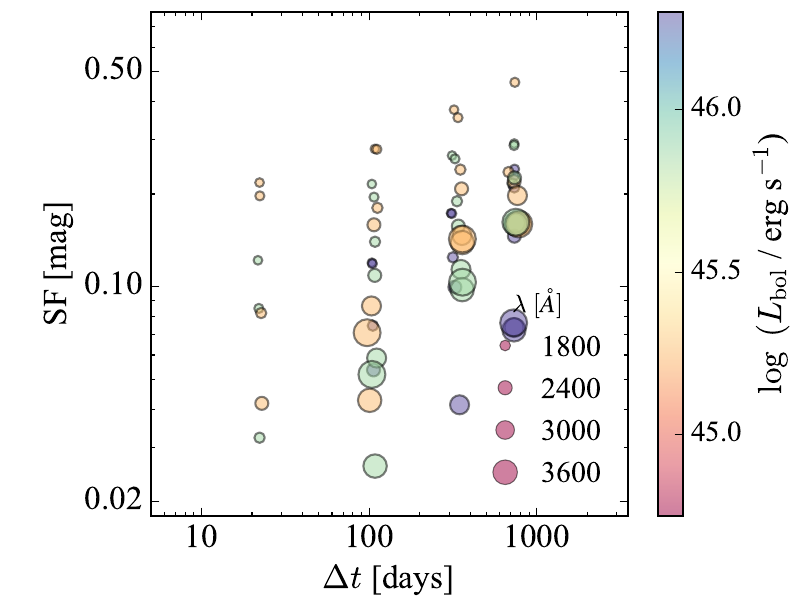}
\caption{
Ensemble SFs of AGNs with a fixed Eddington ratio, grouped by bolometric luminosity.}
\end{figure}

\subsection{Comparison with previous studies}
Mostly, AGN variability has been studied from the photometric dataset obtained with broadband filters \cite[e.g.,][]{macleod_2010,sanchez_2018,arevalo_2023}. Photometric data, however, can be severely affected by the emission from BLR, which can lead to systematic biases in characterizing the AGN variability. For example, \citet{macleod_2010} showed, based on the multi-epoch photometry of QSOs in Stripe 82, that $\tau$ exhibits moderate correlation with BH mass and wavelengths but little correlation with AGN luminosity. More recently, \citet{arevalo_2023} analyzed multi-epoch $g$-band photometry from the Zwicky Transient Facility (ZTF; \citealt{masci_2019}). \cite{burke_2021} analyzed broadband light curves from AGNs across a wide range of BH masses ($10^4 M_\odot < {M_{\rm BH}} \leq 10^{10} M_\odot$), collected from surveys like SDSS and various RM projects. Both studies discovered a correlation between $\tau$ and BH mass, contrary to the results in this study. Although the direct cause of this discrepancy with our findings remains unclear, it is worth noting that, unlike this study, \citet{tang_2023} estimated $\tau$ directly from the SFs or PSDs of multi-year broadband photometric light curves. Accurate measurements of $\tau$ are challenging, even when using light curves with baselines exceeding 10 times the value of $\tau$ (\citealt{kozlowski_2017}).

On the other hand, \citet{tang_2023} analyzed the $\sim 5$-years multi-epoch broadband photometry of $\sim 5,000$ QSOs from the ATLAS survey using the same method in this study and found that $\tau$ is primarily governed by AGN luminosity and wavelength. While this result aligns with our finding, their reported value of $\beta\sim2.4$ is systematically larger than that obtained in this study. To examine the impact of broadband filters, we artificially generated the broadband photometric light curves using the spectroscopic datasets employed in this study. To mimic broadband conditions, we set the width of the continuum windows to 1000 \AA. Applying the same analysis method, we found $\alpha = 0.67\pm0.06$ and $\beta = 2.13\pm0.22$, which agrees well with the results of \citet{tang_2023}. More importantly, excluding continua with central wavelengths longer than 3050 \AA\ yielded a similar trend ($\alpha = 0.68\pm0.07$ and $\beta = 1.77\pm0.28$), with no significant change in the parameters derived from including the entire wavelength range. These findings demonstrate that, in broadband photometric data, contamination from strong emissions and the Balmer continuum originating from the BLR is non-negligible.  
Additionally, the range of AGN luminosity also differs among the observational studies ($\sim10^{46-47}$ erg s$^{-1}$ for \citealt{tang_2023} and $\sim10^{44.8-46.2}$ erg s$^{-1}$ for our study).    

On the other hand, \citet{son_2025} utilized the multi-epoch spectroscopic dataset obtained through the SDSS RM project and analyzed the SFs of $\sim 800$ QSOs, which, to our knowledge, was the first effort to specifically construct the SF for a large sample. Applying the same method in this study, they showed that $\tau$ is mostly determined by AGN luminosity and wavelengths with $\alpha=0.50\pm0.03$ and $\beta=1.42\pm0.09$. It is noteworthy that their results are consistent with ours at the $\sim 2\sigma$ level, despite the differences in the spectroscopic datasets: \citet{son_2025} employed a $\sim 7$-year time series of spectra for hundreds of QSOs, whereas this study used 2 million spectroscopic pairwise combinations for thousands of QSOs. This confirms the reliability of our approach to explore the AGN variability. \citet{son_2025} showed that the fitting result for $\alpha$ and $\beta$ is insensitive to the wavelengths of the continuum windows, revealing that the contribution from the Balmer continuum may not be crucial. However, on the contrary, the best fitting value for $\beta$ changes significantly with and without longer wavelengths above $3050 \AA$. This may be attributed to the fact that our sample comprises less luminous AGNs compared to \citet{son_2025}, and hence the contribution from the host galaxy is more substantial in our sample.     

To mitigate potential biases introduced by broadband photometric data and/or the limited range of AGN physical parameters resulting from a relatively small sample size, simultaneous monitoring with both photometric and spectroscopic datasets over a large sample is ideal. Such an approach will be enabled, for instance, by the Time Domain Extragalactic Survey (\citealt{frohmaier_2025}) conducted with the 4 m Multi-Object Spectroscopic Telescope, in coordination with the Vera C. Rubin Observatory’s Legacy Survey of Space and Time (LSST; \citealt{ivezic_2019}). 

\begin{figure}[t]
\centering
\includegraphics[width=0.45\textwidth, page=1]{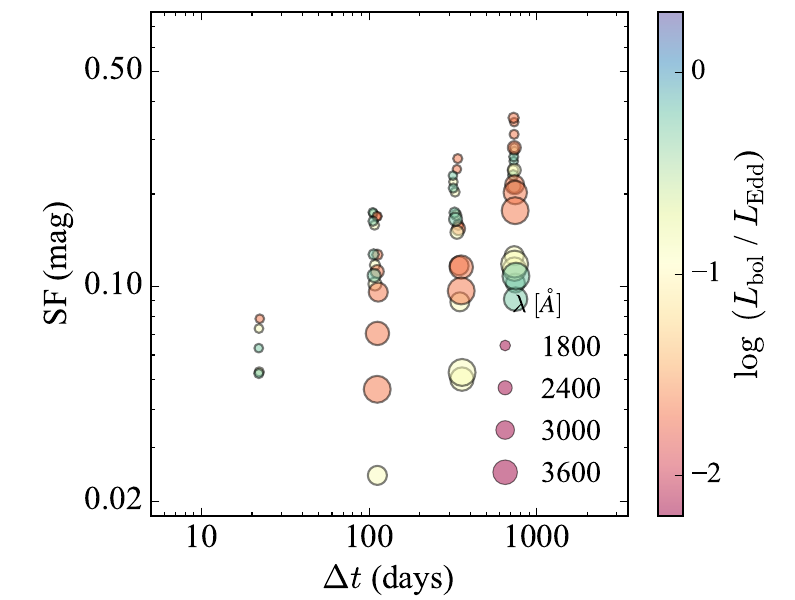}
\caption{
Ensemble SFs of AGNs with a fixed bolometric luminosity, grouped by Eddington ratio.}
\end{figure}

\section{Summary}
Utilizing multi-epoch spectroscopic data of $\sim90,000$ luminous AGNs from SDSS DR16, we examined the continuum ensemble SFs for subsamples categorized by various AGN properties, such as BH mass, AGN luminosity, and Eddington ratio. Based on this analysis, we reached the following conclusions:

\begin{itemize}


\item{Overall, the UV SFs are strongly dependent on wavelength, AGN luminosity, and Eddington ratio. However, they are weakly or barely correlated with the BH mass. To identify the primary parameter determining the SFs of AGNs, we re-examined the SFs by fixing either the AGN luminosity or the Eddington ratio. This revealed that the SFs are solely dependent on wavelength and AGN luminosity.}

\item Under the assumption of a universal SF shared by all the AGNs in which the variability timescale ($\tau$) is determined by wavelength ($\lambda$) and AGN luminosity ($L$), $\tau$ is found to scale with $\alpha=0.62\pm0.07$ and $\beta=1.74\pm0.23$ following the equation $\tau \propto L^\alpha \lambda^\beta$. This finding aligns with the prediction ($\alpha=0.5$ and $\beta=2$) from the standard disk model to within a $2\sigma$ level.  

\item Excluding continua above $3050\AA$ allowed us to minimize the flux contribution from the host galaxy and Balmer continuum. We found $\beta=1.12\pm0.24$, indicating a significant $\sim4\sigma$ deviation from the standard disk model. This outcome may suggest that the temperature profile of the accretion disk is steeper than that predicted by the standard model.

\item The findings of this study significantly deviate from those of previous studies based on broadband photometric data. To robustly examine the dependence of SFs on AGN properties, particularly wavelength, it is essential to use spectroscopic datasets. This underscores the importance of spectroscopic monitoring of AGNs.

\end{itemize}

\begin{acknowledgements}
We are deeply grateful to the anonymous referee for insightful and constructive comments that have substantially enhanced the manuscript.
LCH was supported by the National Science Foundation of China (12233001) and the China Manned Space Program (CMS-CSST-2025-A09). This work was supported by the National Research Foundation of Korea (NRF) grant funded by the Korean government (MSIT) (Nos. RS-2024-00347548 and RS-2025-16066624). 

\end{acknowledgements}

\bibliography{torus}

\end{document}